\documentclass[twocolumn,showpacs,prl]{revtex4}
\usepackage{graphicx}%
\usepackage{dcolumn}
\usepackage{amsmath}
\usepackage{latexsym}
\usepackage{longtable}
\begin{document}
\title{Effect of intrinsic instability of cantilevers on static mode Atomic Force Spectroscopy}
\author{Soma Das\footnote[1]{email:soma@bose.res.in}}
\affiliation{DST Unit for NanoSciences, Department of Materials Science, S.N.Bose National Centre for Basic Sciences, Kolkata  700 098, West Bengal, India.}
\author{P.A. Sreeram\footnote[2]{email:sreeram@iiserkol.ac.in}}
\affiliation{Indian Institute of Science Education and Research, Kolkata  700 106, West Bengal, India.}
\author{A. K. Raychaudhuri\footnote[3]{email:arup@bose.res.in}}
\affiliation{DST Unit for NanoSciences, Department of Materials Science, S.N.Bose National Centre for Basic Sciences, Kolkata  700 098, West Bengal, India.}
\author{Dirk Dietzel\footnote[4]{email:dietzeld@uni-muenster.de}}
\affiliation{Physics Institute, University of M$\ddot u$nster, D-48149 M$\ddot u$nster, Germany and Institute of Nanotechnology, Forschungszentrum Karlsruhe, D-76021, Karlsruhe, Germany.}
\date{\today}
\begin{abstract}
\noindent
%To write
We show that the static force spectroscopy curve is significantly modified due to presence of intrinsic cantilever instability. This instability acts in tandem with such instabilities like water bridge or molecular bond rupture and makes the static force spectroscopy curve (including ``jump-off-contact") dependent on the step-size of the movement of sample stage. A model has been proposed to explain the data. This has been further validated by applying an electric field between tip and substrate which modifies the tip-substrate interaction.
\end{abstract}
\maketitle
Atomic Force Microscopy (AFM) has emerged as a powerful tool, having a wide variety of applications, from understanding the atomic level forces, Casimir force \cite{chen06}, friction at nanometer length scales \cite{evstigneev06,medyanik06} to controlled manipulation of atoms \cite{trevethan07}. This relatively simple instrument has revolutionized our understanding of structures at the nanometer scale and hence the ability to manipulate systems at atomic scales in a wide variety of subjects including material science, soft matter and biology \cite{kuhner06,harris07}.

\begin{figure}
%\vspace{0.01cm}
\begin{center}
\includegraphics[height=5cm]{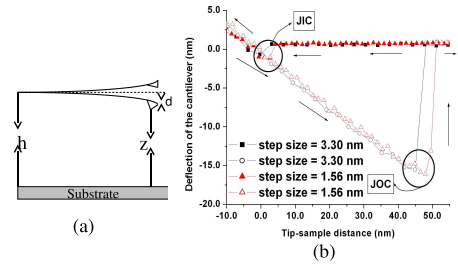}
\end{center}
\vspace{-1.3cm}
\caption{(Colour Online) Schematic diagram of AFM tip and sample assembly (a) and deflection-distance curves at different step sizes taken on Si (b). The dotted line in (a) marks the equilibrium position of the cantilever in the absence of an external force. d is positive when measured upwards. The arrows in (b) show the direction of motion of the cantilever.}
\label{afm1}
\end{figure}

Since the time of its discovery by Binning et. al. \cite{binning82}, many attempts have been made to explain some of the non-intuitive features seen in these systems. For example, the force versus distance (f-h) curves \cite{cappella99} depend on whether the cantilever is approaching towards the sample or retracting away from it (henceforth referred to as the `` approach'' and the ``retract'', respectively), leading to a hysteresis like behaviour as shown in fig.\ref{afm1}. The hysteresis has traditionally been attributed to adhesion due to the layer of water existing on the surface of the sample \cite{cappella97,wen91}, or rupture of molecular bonds \cite{blum01,grobelny06}, and has indeed been used to measure the ``snap off'' force. In this paper we show that the widely used practice of determining the ``snap off" force from the (f-h) curves can be erroneous because intrinsic instability in cantilevers can actually modify the (f-h) curves. We also show how one can properly interpret the (f-h) curves in the context of these instabilities.
In an actual experiment the quantity measured is the cantilever deflection ($d$) as a function of the distance between the sample and the cantilever tip when the tip is in the equilibrium position (in the absence of any external force) ($h$). The force $f = k_c d$, where $k_c$ is cantilever spring constant. It is important to note that although
the steps in which the sample approaches or retracts from the cantilever (the z-controller resolution) is very small ($\approx$ 0.025 \AA{}), the cantilever deflection $d$ is only measured at discrete points in the whole path. In all our discussions below we will define ``step size" ($\delta h$) as the distance between two such neighboring points, and assume that the distance between these points is covered smoothly without any noticable change to the deflection. 
%The important point here is that in the steps intermediate to these points, the cantilever does not relax to a local equilibrium position, while precisely at the points the cantilever does settle down to the local equilibrium. 
If the maximum distance between the cantilever and the surface is $h_{max}$ and the number of data points acquired is $N$, then $h_{max}$ = $ N \delta h$. In all our experiments, $N$ has been kept fixed at 500 (in one direction) and hence $\delta h$ can be varied by varying $h$. 

The two important parameters that one obtains from the experimental f-h curves are the ``jump-into-contact" (JIC) distance obtained from the approach part and the ``jump-off-contact" (JOC) from the retract part. The force, defined by $f^\star=k_c d^\star$, where $d^\star$ is the cantilever deflection at JOC, has traditionally been attributed to adhesion or molecular bond rupture. 
%These forces are effective only if the tip comes in contact with the sample and are independent of $\delta h$.

\begin{figure}
%\vspace{0.1cm}
\begin{center}
\includegraphics[height=5cm]{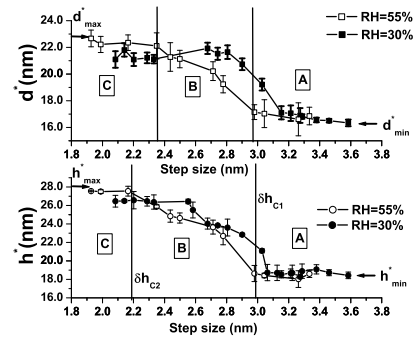}
\end{center}
\vspace{-1.0cm}
\caption{Variation of $d^\ast$ and $h^\ast$ with step size. The closed squares and closed circles show the experimental curves for RH = 30$\%$ and the open squares and the open circles for RH = 55$\%$.}
\label{afm2}
\end{figure}

However, we observe experimentally that $d^{*}$ and $h^{*}$ (the tip-sample separation at JOC) depend on step size ($\delta h$) as shown in fig.~\ref{afm2}. In this paper we show both experimentally and through theoretical analysis that these observed dependencies of $d^{*}$ and $h^{*}$ on $\delta h$ arise due to an intrinsic instability in the cantilever dynamics which manifests itself mainly due to the procedure of data acquisition in most AFM. We show, in particular, that both the instabilities (the intrinsic instability and the ``snap off" instability) occur in tandem. We also find that in UHV-AFM where the ``snap-off" instability is absent, one observes the JIC and JOC arising solely from the intrinsic instability. The intrinsic instability arises mainly due to the motion of the cantilever in a non-linear force field and the two instabilities can be separated out in  a real AFM experiment by acquiring data as a function of $\delta h$. We support our inference by varying the tip-surface force and thus the instability in a controlled manner by applying an electric field between them. We find that the shift of the observed $d^{*}$ and $h^{*}$ on $\delta h$ can be cleanly explained by our model.

\begin{figure}
%\vspace{0.1cm}
\begin{center}
\includegraphics[height=5cm]{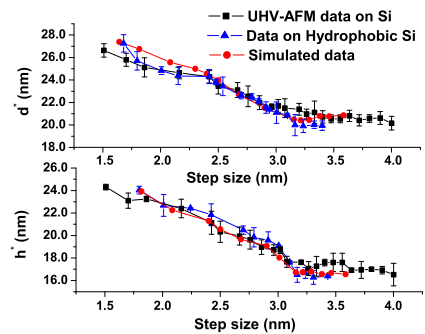}
\end{center}
\vspace{-1.0cm}
\caption{ (Colour Online) Variation of $d^\ast$ and $h^\ast$ with step size. The closed squares show the experimental curves on Si taken in UHV-AFM, the closed triangles show the experimental curves on hydrophobic Si and the closed circles show the simulated curves. The parameters used for the simulation are mentioned in the text.}
\label{afm3}
\end{figure}

The experiments were carried out using a commercial AFM  (Model CP II,Veeco) \cite{veeco} using cantilevers ($k_c \approx 0.1$N/m) with Si$_3$N$_4$ tip on cleaned Si wafers with natural oxide layer on it unless otherwise stated. The cantilever tip had a radius of curvature, $R_t \approx $ 30 nm as determined by direct imaging. For experiment using applied electric field, the substrate was  gold film and a Si cantilever ($k_c$=0.2 N/m) with tip coated with PtIr was used. A d.c. bias was applied to the tip from an external source and the sample was grounded. Experiments were carried out in a glove box with controlled $R_H$ using flow of Ar gas at a temperature controlled environment at 28$^0$ C. The rate of data collection was 0.1 Hz for all the experiments presented here.
Fig.~\ref{afm2} shows a set of $h^\ast$ and $d^\ast$ data plotted as a function of $\delta h$. The data have been obtained from the typical (d-h) curves as shown in fig.~\ref{afm1}. The data taken with two representative humidities ($R_H$=30\% and 55\%) are taken on Si surface with oxide (hydrophilic). Another set of data (fig.~\ref{afm3}) are taken on a hydrophobic surface (created by etching the oxide layer using 50:1 (v/v) HF solution for 30 seconds). All the data show a definite trend. There are three regions in  the data (barring the data taken on the hydrophobic surface). In region A, occurring at highest step size, we find that for $\delta h \geq $  $\delta h_{c1}$, both $h^\ast$ and $d^\ast$, reach a limiting value which is independent of $\delta h$. We call these limiting values $h^\ast_{min}$ and $d^\ast_{min}$, and they are almost independent of the $R_H$ values. In the region C, that occurs for smaller $\delta h$ the $h^\ast$ and $d^\ast$ again reach a limiting value $h^\ast_{max}$ and $d^\ast_{max}$ for $\delta h \leq \delta h_{c2}$ for hydrophilic surface. For $\delta h \leq \delta h_{c2}$ both $h^\ast$ and $d^\ast$ become independent of $\delta h$ and $h^\ast_{max}$, $d^\ast_{max}$ and $\delta h_{c2}$ all depend strongly on $R_H$. In particular $\delta h_{c2}$ is most sensitive to $R_H$ and it increases as $R_H$ is decreased along with the decrease in $h^\ast_{max}$, $d^\ast_{max}$. For the hydrophobic surface (fig.~\ref{afm3}), there is no $\delta h_{c2}$ and $h^\ast$ and $d^\ast$ go on increasing as $\delta h$ is reduced. Data taken in an UHV-AFM is similar to that taken on a hydrophobic surface (there is no $\delta h_{c2}$). The data shown here are representative of a large number of data collected in the controlled experiment. In the region B which is the transition region, both $h^\ast$ and $d^\ast$ increase as $\delta h$ is reduced.
We propose that the two limiting regions in the data (region A and C) are due to the two instabilities that determine the cantilever motion. The instability at lower step size (region C) which depends on the humidity is due to the ``snap-off" phenomena arising from the breaking of the water bridge at the tip-substrate interface. A strong proof in favour of this is the observation that it is absent in the data taken on a hydrophobic surface and in the data taken using UHV-AFM. The instability at higher  step size (region A) is always present and arises due to the intrinsic instability that we describe below. In the following part we use a model proposed earlier by us \cite{das07} to explain our observations. The motion of a cantilever is modeled  by a spring-ball system. The inherent nonlinearity of the cantilever due to its finite dimensions have not been introduced into our calculation, in order to keep things simple. Thus, we write the force balance equation  for the static (or quasi-equilibrium) case as 
\begin{equation}
k_c d = f_{ts}(h+d), 
\label{static_eqn}
\end{equation}
The tip-sample interaction force $f_{ts}(h+d)$ is modeled by a combination of attractive van-der-Waals interaction, for a sphere plate geometry (which is close to the real situation) and the repulsive forces arising due to elastic interaction between the tip and the sample. In order to have a definite result Dejarguin-Muller-Toporov(DMT) \cite{israelachvili91, derjaguin75} force between the tip and the surface has been used. The tip-sample force is thus, formally, given by,
\begin{eqnarray}
\label{force_eqn}
f_{ts}(z) =
\begin{cases} 
- \frac{H R_t}{6 z^2} & for z>a_0,
\\
- \frac{H R_t}{6 a_0^2} + \frac{4}{3}E^\star \sqrt{R_t} (a_0-z)^{3/2} & for
z\leq
a_0.
\end{cases}
\end{eqnarray}
%We have modelled the tip-sample interaction force $f_{ts}(h+d)$ by an attractive van-der-Waals interaction, for a sphere %plate geometry (which is close to the real situation of the AFM experiment) given by,
%\begin{equation}
%\label{force_eqn}
%f_{ts}(h+d) =- \frac{H R_t}{6 (h+d)^2}
%\end{equation}
where $H$ and $R_t$ are the Hamaker constant and the radius of curvature of the tip respectively. The attractive force is the only force present when $h+d >$ the intermolecular distance ($a_0$), whereas when $h+d <$ $a_0$ the force has a repulsive component, which incerases with reducing $h$. The repulsive component typically ensures that $h+d$ $>$ 0. It is interesting to note that, while the repulsive force is essential, the qualitative understanding of the f-h curves, comes even when the repulsive force is taken to be absent. In \cite{das07} we have ignored the repulsive interaction for obtaining the exact solutions to the equation of motion of the cantilever. This will produce a slight deviation from the actual results, however, this will not change the conclusion. %However, the intuitive understanding of the exact solution for the equations, obtained by %ignoring the repulsive part, is very attractive and hence we will follow the same.  
%In our analysis, however, the form of the repulsive force is unimportant since the hysteresis behaviour occurs typically %at distance much larger than $a_0$.

%\begin{eqnarray}
%\label{force_eqn}
%f_{ts}(z) =
%\begin{cases} 
%- \frac{H R_t}{6 z^2} & for z>a_0,
%\end{cases}
%\end{eqnarray}
%Here, $z=h+d$, $a_0$ is an intermolecular distance, $H$ is the Hamaker constant. 
%The form of the  force is chosen for a sphere-plate geometry, which is close to the real situation in an AFM experiment.
% Since we are working in the region $z > a_0$ where the JOC occurs,  there is no contact of the cantilever with the
%substrate. Thus  we do not include the contact force due to elastic deformation of the surface. 
From eq.~\ref{static_eqn} and eq.~\ref{force_eqn} (in the region $h+d > a_0$), we obtain the equation for the deflection ($d$) as,
\begin{equation}
\tilde d(1+\tilde d)^2+\tilde a = 0.
\label{mod_static}
\end{equation}
where $\tilde d=d/h$ and $\tilde a=HR_t/6k_ch^3$ are dimensionless. The three exact solutions of this equation are already given in \cite{das07}, therefore, we are not mentioning it here. For further discussion, we will refer these three solutions as $\tilde d_1$, $\tilde d_2$ and $\tilde d_3$. 
%\begin{eqnarray}
%\tilde d_1 &=& -\frac{2}{3} + (S+T) \nonumber \\
%\tilde d_{2,3} &=& -\frac{2}{3} -\frac{1}{2}(S+T)\pm \frac{\sqrt{3}}{2}\imath(S-T)
%\nonumber \\
%\tilde d_3 &=& -\frac{b_2}{3} -\frac{1}{2}(S+T)-\frac{\sqrt{3}}{2}\imath(S-T)
%\label{solutions}
%\end{eqnarray}
%where, $S = \left(R + \sqrt{D}\right)^{1/3}$, $T = \left(R - \sqrt{D}\right)^{1/3}$, $R = %(2-27 \tilde a)/54$, $Q = -1/9$ and $D = Q^3 + R^2$.
\begin{figure}
%\vspace{0.1cm}
\begin{center}
\includegraphics[height=5cm]{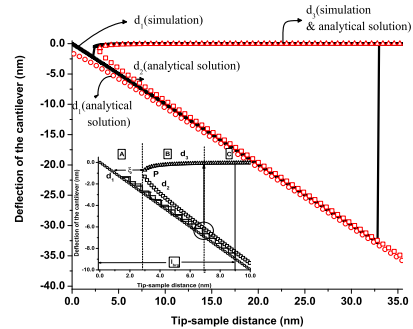}
\end{center}
\vspace{-1.0cm}
\caption{(Colour Online) Plot of the simulated d-h curves (black line) and the three analytical solutions (red symbols) of eq.~\ref{mod_static} as a function of tip-sample distance (h) for the parameters mentioned in the text. For analytical solutions, the open circles ($d_1$=$\tilde d_1.h$) and open triangles ($d_3$=$\tilde d_3.h)$ represent stable solutions and the open squares ($d_2$=$\tilde d_2.h$) represent the unstable solution. Three different regions (A,B and C) are shown in the inset.}
\label{afm4}
\end{figure}
Solution $\tilde d_1$ is real while $\tilde d_2$ and $\tilde d_3$ are either both real or complex conjugate of each other, depending on the parameters of the equation. Fig.~\ref{afm4} shows the simulated d-h curves (approach and retract) in presence of both attractive and repulsive part of the tip-sample interactions and also the analytical solutions of eq.~\ref{mod_static} as a function of the tip-sample distance ($h$) for $HR_t$=2.2 x 10$^{-27}$ N.m$^2$ (appropriate for our experimental conditions) and $a_0$=0.172 nm. 
%In fig.~\ref{afm4} the open symbols represent the solutions in absence of electric field %whereas the closed symbols represent the solutions in presence of electric field. In %presence of electric field, the three solutions are represented by $(d_1)^{'}$, $(d_2)^{'}$ %and $(d_3)^{'}$. 
In simulated d-h curves we get realistic $d_1$ (= $\tilde d_1.h$) as in this case both attractive and repulsive part of the tip-sample interactions are present. One can find out from fig.~\ref{afm4} that $d_1+h$, obtained from simulation, is always positive whereas in case of analytical solution it is negative because the repulsive part of the tip-sample interaction has not been considered. It can be noted that the JIC position also matches quite well in these two cases. The solutions corresponding to $d_2$ and $d_3$ approach each other and they meet at a point P (for example, at $h$ $\approx$ 3.0 nm) and as $h$ is reduced below this point they become complex. The distance of point P from the substrate ($\xi$) is the JIC point, which arises from the intrinsic cantilever instability \cite{das07}. 
%Stability of the solutions have been tested finding the sign of the derivative of Eq.(\ref{static_eqn}) with respect to $d$ at each value of $h$. 
Here $\tilde d_1$ and $\tilde d_3$ are stable solutions, while $\tilde d_2$ is unstable. Hence, the tip will either equilibriate to the solution $\tilde d_1$ or $\tilde d_3$.

We use fig.~\ref{afm4} to explain the observed data. In most commercial AFM, during the process of data acquisition for the d-h curves, the motion of the cantilever is quasi-continuous, i.e., at each point the initial deflection ($d$) of the cantilever is determined by its final deflection at the previous point. In fig. \ref{afm4} (in the inset), we show two examples of the paths traced by the cantilever (shown as steps). In one case (solid line steps), for relatively large step sizes
($\delta h_{c2} < \delta h < \delta h_{c1}$), the intrinsic instability dominates, and the jump from $\tilde d_1$ to $\tilde d_3$ occurs when the equilibrium position at the point just prior to the jump, takes the cantilever across $\tilde d_2$ (marked by a circle in fig. \ref{afm4}). In the other case (dotted line steps), for relatively smaller step sizes ($\delta h \sim \delta h_{c2}$), the ``snap off'' instability dominates and causes a jump across the solution $\tilde d_2$. Here, $l_{brg}$ determines the scale at which the water bridge snaps off, causing a jump across the solution $\tilde d_2$. If $\delta h\geq \xi$, then, during the retract part, the cantilever tip will jump directly to the stable solution 
$\tilde d_3$ and $d^{*}$ and $h^{*}$  both become essentially independent of $\delta h$. This corresponds to the region A, where the intrinsic instability is solely responsible for the JOC and which among other things depends on $k_c$, Hamaker constant $H$ and tip radius $R_t$. In region A we can thus identify  $\xi \approx \delta h_{c1}$. 
%Depending on  the substrate hydrophilicity and the $R_H$, one may have a water bridge between the tip and the substrate. 
%When the step size is reduced from  $\delta h_{c1}$
 %the tip will follow the trajectory as shown in fig.~\ref{afm4}. The quasi-continuous nature of the tip
 %motion is marked as the steps. 
In region B both $h^\ast$ and $d^{*}$ increase as $\delta h$ is decreased. This is the region described  above (solid line steps in fig. \ref{afm4}). In absence of ``snap off'' instability, the region B extends all the way down to very small step sizes, as seen on the experiments on hydrophobic surfaces.
On  the other hand, if the ``snap off'' instability is present and $\delta h \sim \delta h_{c2}$, the JOC occurs when $h \sim l_{brg}$, as discussed above (dotted line steps in fig. \ref{afm4}). This is the region we identify as region C. In this region $h^{*}$ and $d^{*}$ are independent of $\delta h$ and $l_{brg}$ is dependent only on $R_H$ and $R_t$. 
%this case as $\delta h$ decreases, $h^\ast$ increases.
%If the distance travelled $h$ is smaller than the critical length ($l_{brg}$) at which the water bridge snaps, the tip 
%will jump back to the solution $\tilde d_3$, when its trajectory hits the unstable solution $\tilde d_2$.
%This will determine the $h^{*}$ for the given $\delta h$ as well as the cantilever deflection $d^\ast \approx h^\ast\times (\tilde d_{3}- \tilde d_{1})$, with both $\tilde d_1$ and $\tilde d_3$ evaluated at $h^\ast$. 
%This is the region B, where both $h^\ast$ and $d^{*}$ increase as $\delta h$ is decreased.
 %However, if the tip trajectory is such that $h$ becomes longer than $l_{brg}$, the cantilever will become unstable and
 %will jump back to its rest position after snapping off the bond. In this region $h \approx h^{*}\approx l_{brg}$ and even
 %if we reduce $\delta h$ further the values of $h^{*}$ and $d^{*}$ will not change. This is the region C and we identify
 %the particular value of $\delta h$ for which this occurs as the $\delta h_{c2}$ , as observed experimentally. In this 
% case $h^\ast_{max}\approx l_{brg}$ , $d^\ast_{max}$ will again be determined by the value of ($\tilde d_{3}- \tilde
% d_{1}$) evaluated at appropriate $h$, as marked in the figure. 
Thus the qualitative discussion based on fig.~\ref{afm4} clearly identifies the regions of the observed curve and the instabilities that give rise to them. Thus in an actual experiment the (d-h) curves need to be taken as a function of $\delta h$ and the regions corresponding to the two instabilities can be clearly identified. In fig.~\ref{afm3}, we show the data taken in an UHV-AFM, on a hydrophobic surface (which show only intrinsic cantilever instability) and the actual calculated curve based on the method above. The data for the curves taken with water bridges can also be fitted by taking $h^\ast_{max}$ as an adjustable parameter. It can be seen that the model proposed by us can explain the data in all the three regions. 

In this investigation, the ``snap-off" occurs due to the instability of the water bridge that forms between the tip and the sample. It is shown earlier \cite{andrienko04} that, for a sphere-plate geometry, depending on the radius of curvature of the tip, the water bridge configuration becomes metastable for a particular sphere-sample separation when  $\frac{R_t}{h}$ becomes $\sim$ 1.0, where $R_t$ is the radius of curvature of the tip. For the tip used $R_t$=30 nm, this should happen for $h \sim 30$ nm which matches very well with the value of $h^\ast_{max}\approx 26.5$ nm observed experimentally.
\begin{figure}
%\vspace{0.1cm}
\begin{center}
\includegraphics[height=4cm]{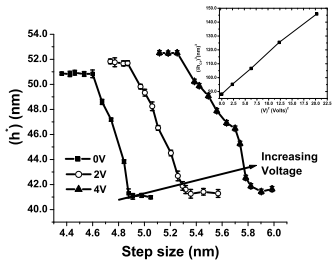}
\end{center}
\vspace{-1.0cm}
\caption{Variation of $h^\ast$ with step size as a function of bias voltage. The closed squares show the curve for V=0V, the open circles for V=2V and the closed triangles for V=4V. Variation of $\delta h_{c1}$ as a function of applied voltage (V) is shown in the inset.}
\label{afm5}
\end{figure}
The main proposal of the paper that there is an intrinsic instability of the cantilever can be further tested if we can modify the $f_{ts}$ in a controlled way. In region A, $\xi \approx \delta h_{c1}$ and our model gives $\xi\approx (\frac{1.12HR_t} {k_c })^{1/3}$. From the experimentally determined $H$ and $R_{t}$ we find that calculated $\delta h_{c1}\approx 2.9$ nm and experimentally obtained value is $\approx 3.0$ nm. We have also checked that if we use softer cantilever having $k_c$ = 0.03 N/m ($R_t$ = 25 nm) then $\delta h_{c1}$ shifts to 4.32 nm (calculated value of $\delta h_{c1}$ in this case is 4.29 nm).
 
To establish the validity of our hypothesis we used electric field to control $f_{ts}$ and obtained the (d-h) curves with applied electric field. The force due to the applied electric field \cite{hudlet98} adds to the force term due to the van der Waals force and will give rise to an effective Hamaker constant $H_{eff}$=$H$+$C.V^{2}(C \approx Constant)$. The enhanced $H_{eff}$ will make $\xi$ larger and will also shift $\delta h_{c1}$ to a higher value. This can be clearly seen in fig.~\ref{afm5}, where we show the data taken in electric field. It has been observed that $\delta h_{c1}$ clearly shifts to higher value with a small applied field. Since $\delta h_{c1}\approx\xi \approx (\frac{1.12HR_t} {k_c })^{1/3}$, in the applied electric field one would observe that $\delta h_{c1}\sim H_{eff}^{1/3}\sim V^{2/3}$. A plot of ${\delta h_{c1}}^3$ vs $V^2$ in the inset of fig.~\ref{afm5} shows that this dependence indeed exists. We also observe a shift of $\delta h_{c2}$, $h^\ast_{max}$  and $d^\ast_{max}$ to higher values on application of electric field. This instability is related to water bridges \cite{sacha06}. It has been seen recently that water bridges become more stable in an applied electric field. The stability of the water bridge will be reflected in enhancement of $l_{brg}$ in an applied electric field leading to enhancement of the attractive force as shown in \cite{sacha06}. 

In summary, we have shown that the static d-h curves for an AFM, depends on the intrinsic instability of the microcantilever of the AFM. 
%We have shown that the calculation of the ``snap off'' force can be done correctly only at %small step sizes ($ \delta h \le \delta h_{c2}$). 
The phenomena like JIC and JOC occur even in absence of water bridge snap-off as in an UHV-AFM and on hydrophobic surface. At larger step sizes, the intrinsic instability dominates over the ``snap off'' instability, leading to erroneous results in the calculation of these forces. The instabilities due to ``snap-off" forces dominate at smaller step sizes.
%Using a model, we have shown that the ``Jump-off-contact" phenomena observed in the deflection-distance curves is related to the intrinsic instability. 
We have also shown experimentally that the intrinsic instability due to cantilever can be controlled by an applied electric field.\\ 
%The theory has been verified recently using Ultra High Vacuum AFM also, details of which will be presented elsewhere.\\
The authors want to thank the Department of Science and Technology, Government of India for financial support as a Unit for Nanoscience. Authors thank Prof. H. Fuchs for allowing use of UHV-AFM of data collection.


\begin{thebibliography}{0}
\bibitem{chen06}  F. Chen, G. L. Klimchitskaya, V. M. Mostepanenko and U. Mohideen, Phys. Rev. Lett. {\bf 97} 170402 (2006).
\bibitem{evstigneev06}  Mykhaylo Evstigneev, Andr\'{e} Schirmeisen, Lars Jansen, Harald Fuchs and 
Peter Reimann, Phys. Rev. Lett. {\bf 97} 240601 (2006).
\bibitem{medyanik06}  Sergey N. Medyanik, Wing Kam Liu, In-Ha Sung and Robert W. Carpick, Phys. Rev. Lett. {\bf 97} 136106 (2006).
\bibitem{trevethan07}  T. Trevethan, M. Watkins, L. N. Kantorovich and A. L. Shluger, Phys. Rev. Lett. {\bf 98} 28101 (2007).
\bibitem{kuhner06}  Ferdinand K\"uhner, Matthias Erdmann and Hermann E. Gaub, Phys. Rev. Lett. {\bf 97} 218301 (2006).
\bibitem{harris07}  Nolan C. Harris, Yang Song, and Ching-Hwa Kiang, Phys. Rev. Lett. {\bf 99} 68101 (2007). 
\bibitem{binning82} G. Binning, H. Rohrer, Ch. Gerber, and E. Weibel, Phys. Rev. Lett. {\bf 49} 57 (1982).
\bibitem{cappella99} B. Cappella  and G. Dietler, Surface Science Reports {\bf 34}, 1 (1999).
%\bibitem{ref2} H.J. Butt, B. Cappella and M. Kappl, Surface Science Reports {\bf 59}, 1 (2005).
\bibitem{cappella97} B. Cappella, P. Baschieri, C. Frediani, P. Miccoli and C. Ascoli, Engineering in Medicine and Biology Magazine IEEE {\bf 16}, 58 (1997).
\bibitem{wen91} H. H. Wen, A.M. Bar\'{o} and J.J. S\'{a}enz, Jour. Vac. Sci. Tech.B {\bf 9}, 1323 (1991).
\bibitem {blum01} M. He, A.S. Blum, D.E. Aston, C. Buenviaje, R.M. Overney and R. Luginb\"uhl, Journal Of Chemical Physics {\bf 114}, 1355 (2001).
\bibitem {grobelny06} J. Grobelny, N. Pradeep, D.-I. Kim and Z.C. Ying, Applied Physics Letters {\bf 88}, 091906 (2006).
%\bibitem {ref7} A. Marmur, Langmuir {\bf 9}, 1922 (1993).
\bibitem {veeco} Veeco Instruments Inc. Corporate Headquarters 100 Sunnyside Blvd. Ste. B Woodbury New York 11797-2902.
\bibitem {das07} S. Das, P.A. Sreeram and A.K. Raychaudhuri, Nanotechnology {\bf 18}, 035501 (2007).
\bibitem {israelachvili91} J. Israelachvili, Intermolecular and Surface Forces 
(Academic Press, London) (1991).
\bibitem{derjaguin75} Derjaguin B V, Muller V M and Toporov Y P, Jour. Colloid Interface Sci., {\bf 53}, 314 (1975).
\bibitem {andrienko04} D. Andrienko, P.Patr\'{i}cio and O.I. Vinogradova, Journal Of Chemical Physics {\bf 121}, 4414 (2004).
\bibitem {hudlet98} S. Hudlet, M. Saint Jean, C. Guthmann, and J. Berger, Eur. Phys. J. B {\bf 2}, 5 (1998).
\bibitem {sacha06} G.M. Sacha, A. Verdaguer, and M. Salmeron, J. Phys. Chem. B {\bf 110}, 14870 (2006); G.M. Sacha, J. J. S\'{a}enz, M. Calleja and R. Garc\'{\i}a, Phys. Rev. Lett. {\bf 91} 056101-1 (2003).
%Sacha G\'{o}mez-Mo\~{n}ivas, Juan Jos\'{e} S\'{a}enz, Montserrat Calleja and Ricardo Garc\'{\i}a, Phys. Rev. Lett. {\bf 91} 056101-1 (2003). 
\end{thebibliography}
\end{document}